\documentclass[conference]{IEEEtran}
\makeatletter
\def\ps@headings{%
\def\@oddhead{\mbox{}\scriptsize\rightmark \hfil \thepage}%
\def\@evenhead{\scriptsize\thepage \hfil \leftmark\mbox{}}%
\def\@oddfoot{}%
\def\@evenfoot{}}
\makeatother
\newtheorem{lem}{Lemma}
\newtheorem{thm}{Theorem}

\newtheorem{defn}{Definition}

\newcommand{\probb}{\mathds{P}}

\newcommand{\B}{\mathcal{B}}
\newcommand{\real}{\mathds{R}}
\newcommand{\E}{\mathcal{E}}
\newcommand{\C}{\mathcal{C}}
\newcommand{\D}{\mathcal{D}}
\newcommand{\T}{\mathcal{T}}
\newcommand{\U}{\mathcal{U}}
\newcommand{\V}{\mathcal{V}}
\newcommand{\M}{\mathcal{M}}
\newcommand{\I}{\mathcal{I}}
\newcommand{\order}{\mathcal{O}}
\newcommand{\Q}{\mathcal{Q}}
\newcommand{\ess}{\mathcal{S}}
\newcommand{\W}{\mathcal{W}}
\newcommand{\ind}{\mathds{1}}

\usepackage{amssymb, amsmath, amstext, color, epsfig, dsfont, wrapfig, url}

\title{Polynomial-complexity, Low-delay Scheduling for SCFDMA-based Wireless Uplink Networks (Technical Report)}
\author{Shreeshankar Bodas, Bilal Sadiq \\ Qualcomm, Inc., Bridgewater, NJ 08807
}

\begin{document}

\maketitle

{\abstract Uplink scheduling/resource allocation under the single-carrier FDMA constraint is investigated, taking into account the queuing dynamics at the transmitters. Under the single-carrier constraint, the problem of MaxWeight scheduling, as well as that of determining if a given number of packets can be served from all the users, are shown to be NP-complete. Finally, a matching-based scheduling algorithm is presented that requires only a polynomial number of computations per timeslot, and in the case of a system with large bandwidth and user population, provably provides a good delay (small-queue) performance, even under the single-carrier constraint.

In summary, the results in first part of the paper support the recent push to remove SCFDMA from the Standards, whereas those in the second part present a way of working around the single-carrier constraint if it remains in the Standards.}

{\keywords Uplink scheduling, single-carrier FDMA, Batch-and-allocate}

\section{Introduction}
\label{sec:intro}

In the recent years, we have witnessed an explosion in the numbers and capabilities of hand-held wireless communication devices, and consequently their data consumption. Real-time, i.e., delay-constrained data traffic (voice/video/gaming/$\dots$) constitutes a significant fraction of the overall over-the-air data demand. The demand for high-quality data, and in large quantities, is ever-growing, but the wireless resources are not growing nearly as fast. It is therefore important to design efficient methods of sharing the resources across multiple users in order to guarantee a good quality of service. In this paper, we focus on the problem of resource allocation on the uplink (user to base-station) of wireless networks.

The 3GPP LTE (Long-Term Evolution) standard has chosen the single-carrier frequency division multiple access (SCFDMA) technology as the uplink multiple access technology~\cite{3GPP_LTE}. The SCFDMA can be thought of as a special case of the orthogonal frequency division multiple access (OFDMA) technology used for the downlink of 3GPP LTE. In OFDMA, the available bandwidth at the base-station is partitioned into a number of orthogonal frequency sub-bands, and a given user can be allocated any subset of the frequency sub-bands for his/her downlink traffic under the condition that a given frequency sub-band can be allocated at most one user. In SCFDMA, there is an additional constraint that a given user can be allocated only \emph{consecutive} frequency sub-bands. For example, consider a system with $2$ users $x, y$ and $3$ frequency sub-bands $f_1, f_2, f_3.$ Then $(x, f_1), (x, f_2), (y, f_3)$ is a valid SCFDMA allocation, while $(x, f_1), (x, f_3), (y, f_2)$ is not. We refer to this additional constraint as the single-carrier constraint. The main reason for the choice of SCFDMA for the uplink is that it results in a lower PAPR (peak-to-average power ratio) than OFDMA.

In this paper, we show that the single-carrier constraint alone is enough to make certain scheduling problems hard (formally, NP-complete). The classic MaxWeight scheduler~\cite{TasEph_92} is throughput-optimal for the uplink network under very mild assumptions on the arrival and channel processes (see~\cite{ErySriPer_04}), but selecting a weight-maximizing schedule is NP-complete (Theorem~\ref{mw_hard}). Another natural, myopic, ``greedy'' scheduler for the scheduling problem described in Section~\ref{sec:sysmodel} operates as follows: given a queue-length vector and a matrix of the rates at which the frequency sub-bands can serve the individual user-queues, does there exist an allocation that serves $x_i$ packets from the user-queue $Q_i$? This scheduler is interesting because by choosing appropriate values of $x_i$s in every scheduling period, the per-user queues can be kept small. For example, the values of $x_i$ can be chosen to equalize the queue-lengths after service. For the downlink scheduling problem, in absence of the single-carrier constraint, this scheduler is shown to have good delay properties~\cite{srb_phd_thesis_10}; but under the single-carrier constraint, implementing it requires solving an NP-complete problem (Theorem~\ref{pd_hard}).

In the light of these negative results, we focus on a simple, i.i.d. arrival and channel model, and design an algorithm called Batch-and-allocate (BA) scheduler as the main contribution of this paper. This scheduler results in a good delay (small-queue) performance for the system, and can be implemented in polynomial number of computations per timeslot, \emph{even under the single-carrier constraint.}

The qualitative messages from the paper are: (i)~The single-carrier constraint, while attractive from a power amplifier point of view, severely restricts the class of possible scheduling policies. There has been a recent push to remove it from the standards (e.g., clustered SCFDMA~\cite{3GPP_LTE_advanced, rws_120022}) and this paper can be seen as an argument in its favor. (ii)~Although the uplink scheduling problem is intractable under the single carrier constraint, we can guarantee a good quality of service for ``regular'' arrival and channel processes, \textit{if the system has a large number of users and proportionally large bandwidth.}

\section{Related Work}
\label{sec:rel_wk}

Scheduling and resource allocation for the wireless uplink network is a well-investigated problem. Researchers have studied this problem from the point of view of maximizing a system-wide utility function~\cite{HuaSubAgrBer_09, RenStoVis_10, MadRay_11}, orderwise delay-optimal scheduling~\cite{neely_08}, successive interference cancellation to allow for simultaneous transmissions from users~\cite{MolGha_11}, and so on. A majority of the previous work on the problem either does not consider the single-carrier constraint, or allows for fractional server (i.e., frequency sub-band) allocation, thus circumventing the inherently discrete nature of the allocation problem. In wireless uplink systems where frequency sub-bands are grouped together, the fractional server allocation is a reasonable assumption. A recurring theme in the prior work is to initially ignore the single-carrier constraint, come up with an allocation of the frequency sub-bands to the users that optimizes a certain objective, and then use heuristics to modify that allocation to incorporate the single-carrier constraint. This approach usually leads to a loss of performance. In contrast, in this paper, we strictly adhere to the single-carrier constraint even in the algorithm design part, and do not perform any fractional server allocations. We present an algorithm that is designed with the single-carrier constraint in mind, and which yields a good small-buffer performance under a variety of changes to the basic system model. To the best of our knowledge, this is the first characterization of the small-queue performance of the uplink network in the large- system limit.

\section{System Model}
\label{sec:sysmodel}

\begin{wrapfigure}{r}{2.0in}
	\centering
	\resizebox{2.0in}{!}{\input{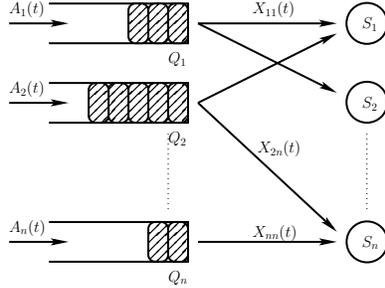}}
	\caption{System Model}
	\label{fig:sysmodel}
\end{wrapfigure}

We consider a discrete-time queuing system with $n$ queues and $n$ servers, as shown in Figure~\ref{fig:sysmodel}.
Here the $n$ queues represent the packet queues at the $n$ uplink transmitters, and the $n$ servers represent the $n$ orthogonal uplink frequency sub-bands. The queues can store any number of packets until they are served, so that there are no dropped packets. Table~\ref{tbl:notation} summarizes the notation used throughout this paper.
\begin{table}[hbcp]
\begin{tabular}{lp{2.8in}}
$\Q$ & The set of $n$ queues $\{Q_1, \dots, Q_n\}$\\
$\ess$ & The set of $n$ servers $\{S_1, \dots, S_n\}$\\
$Q_i(t)$ & The length of $Q_i$ at the end of timeslot $t$\\
$\hat{Q}(t)$ & $\max \{Q_i(t): 1 \le i \le n\}$\\
$X_{ij}(t)$ & The number of packets that the server $S_j$ can potentially serve from $Q_i$ in timeslot $t$\\
$A_i(t)$ & The number of arrivals to $Q_i$ at the beginning of timeslot $t$\\
$[n]$ & The set $\{i : 1 \le i \le n\}$\\
$a^+$ & $\max(a, 0)$\\
$|A|$ & The cardinality of set $A$\\
$\real_+$ & The set $[0, \infty)$ of nonnegative real numbers\\
$\Delta_k$ & The probability simplex in $\real^{k}$
\end{tabular}
\caption{Notation}
\label{tbl:notation}
\end{table}

\textbf{Arrival and channel processes:} We assume that the arrivals to the queues and the channel realizations are i.i.d across queues, servers, and timeslots. More precisely,
\begin{enumerate}
	\item The number of arrivals to $Q_i$ at the beginning of timeslot $t$ are i.i.d. across timeslots and queues, and obey $\probb(A_i(t) = m) = p_m$ for $0 \le m \le M,$ $p_i > 0$ for all $i,$ and $\sum_{i = 0}^M p_i = 1.$
	\item The number of packets that the server $S_j$ can serve from $Q_i$ in timeslot $t$ are i.i.d across queues, servers and timeslots, and obey $\probb(X_{ij}(t) = k) = q_k$ for $0 \le k \le K,$ $q_i \ge 0$ for all $i,$ and $\sum_{i = 0}^K q_i = 1.$
	\item\label{load_rf_nonzero} There exists $\alpha \in (0, 1)$ such that $\sum\limits_{i = 0}^M p_i \left\lceil \dfrac{i}{K} \right\rceil = 1 - \alpha.$
\end{enumerate}

We make the assumption $p_i > 0$ for all $0 \le i \le M$ only to avoid trivialities; our results or proof techniques are in no way dependent upon this assumption. We also assume that $M > K,$ since otherwise, allocating just one server (with highest supported rate $K$) is enough to serve all the new arrivals to a queue in a given timeslot, and the single-carrier constraint in the problem can be easily circumvented by the matching-based algorithms for the downlink, such as those in~\cite{BodShaYinSri_sig_09}. Our objective is to define a service policy, quantified by the random variables $Y_{ij}(t) \in \{0, 1\}$ for $i, j \in [n]$ and for all $t,$ where $Y_{ij}(t) = 1$ if the server $S_j$ serves the queue $Q_i$ in timeslot $t,$ and $0$ otherwise. The random variables $Y_{ij}(t)$ are allowed to depend upon the entire past of the system and the arrivals and channel realizations in the (current) timeslot $t,$ but are required to satisfy the following conditions:
\begin{enumerate}
	\item $\sum_{i = 1}^n Y_{ij}(t) \le 1$ for all $i, j, t.$
	\item If $Y_{ir}(t) = Y_{is}(t) = 1$ for some $1 \le r < s \le n,$ then $Y_{ij}(t) = 1$ for all $r < j < s,$ all $i \in [n].$
\end{enumerate}
The first condition implies that a given server can serve at most one queue in any timeslot. The second condition models the single-carrier constraint. The queues evolve according to
\begin{equation}
Q_i(t) = \Big(Q_i(t - 1) + A_i(t) - \sum_{j = 1}^n X_{ij}(t) Y_{ij}(t)\Big)^+.
\label{eq:queue_evol}
\end{equation}
Our objective is to define a scheduling policy that, for every integer $b \ge 0,$ results in a strictly positive value of
\[\I(b) := \liminf_{n \to \infty} \frac{-1}{n} \log \probb\left(\max_{1 \le i \le n}Q_i(t) > b\right),\]
where $\probb(\cdot)$ refers to the stationary distribution of the queue-length process. The function $\I(\cdot)$ is called the rate-function in large deviations theory~\cite{dembozeitouni}. In order to guarantee a good small-queue performance, our true objective is to minimize the ``overflow'' probability, i.e., the probability of the event $\{\max_{1 \le i \le n} Q_i(t) > b\}.$ In real systems with a large number of users and proportionally large bandwidth, the rate-function maximization is a useful and reasonable surrogate for this objective. If $\I(b) > 0,$ then the probability of the overflow event rapidly diminishes to $0$ with the system-size. Hence in this paper, we focus on policies that result in a strictly positive value of the rate-function. The assumption~\ref{load_rf_nonzero} is a necessary condition for the rate function to be nonzero, even without the single-carrier constraint~\cite{BodShaYinSri_infocom_11}. \textit{Our main contribution is an algorithm that yields a positive value of the rate-function under this assumption.}

Note: In the rest of the paper, for simplifying notation, we make statements like ``allocate $n/2$ servers to a queue.'' What we actually mean is the integer part (or floor) of the corresponding fraction. We \emph{never} make fractional server allocations. We are interested in the large deviations results ($n$ large). In this regime, the rounding has no effect on the analysis. We do not discuss this issue further in this paper.

\section{Computational Hardness}
\label{sec:hardness}

In this section, we establish that in the presence of the single-carrier constraint, certain (otherwise simple and interesting) scheduling policies are NP-complete. We use a construction almost identical to the one from~\cite{LeePefMeyXuLu_09}. In~\cite{LeePefMeyXuLu_09}, the authors establish the NP-hardness of the single-carrier scheduling problem in the context of proportionally fair (PF) scheduling. Their reduction can be modified to suit in our case. The reasons that we provide a detailed account here, as opposed to merely citing their result, are: (i)~their result is not directly applicable in our case: it is concerned with PF scheduling, and (ii)~their construction is cryptic to the authors of this paper, with a number of key proof details missing.

In the multi-queue multi-server setup described here, a natural, myopic way to minimize the probability that the longest queue exceeds a given constant $b$ is to select, in every timeslot, that allocation of the servers to the queues that minimizes the maximum queue-length. This requires answering the question: \emph{can a queue $Q_i$ be allocated at least $w_i$ units of service, $i \in [n]$?} A simpler question as defined in Definition~\ref{defn:pkt_drain} is: can a total of $W$ packets be drained from the queues? Our objective is to show that even this simpler problem is NP-complete under the single-carrier constraint.

{\defn[Packet-draining problem (PD)]\label{defn:pkt_drain} Consider a queue-length vector $[Q_1, \dots, Q_k]$ and a set of servers $\{S_1, \dots, S_m\},$ where the server $S_j$ can serve $X_{ij}$ packets from the queue $Q_i.$ A finite integer $W \ge 0$ is given. Determine if, under the single-carrier allocation constraint, there exists an allocation of the servers to the queues that serves a total of at least $W$ packets. \hfill $\diamond$}

{\thm\label{pd_hard} The packet-draining problem (PD) is NP-complete.}
\begin{proof}
Please see Appendix~\ref{pd_hard_proof}.
\end{proof}

We now focus on the problem of MaxWeight scheduling under the single-carrier constraint. This classic scheduling algorithm was introduced in~\cite{TasEph_92} and is known to be throughput-optimal (i.e., makes the queue-length Markov chain positive recurrent if there is any other algorithm that can do so) in a variety of situations, including under the single-carrier constraint, even under more general (e.g., correlated) arrival and channel processes~\cite{ErySriPer_04}. But as is established next, implementing it is computationally intractable unless P=NP.

{\defn[MaxWeight problem (PM)]\label{defn:mw_prob} Consider a set of queues $[Q_1, \dots, Q_k]$ with lengths $[L_1, \dots, L_k],$ and a set of servers $\{S_1, \dots, S_m\},$ where the server $S_j$ can serve $X_{ij}$ packets from the queue $Q_i.$ A finite integer $W \ge 0$ is given. Let $Y_{ij} = 1$ if the server $S_j$ is allocated to $Q_i,$ and $0$ otherwise. Determine if, under the single-carrier allocation constraint, there exists an allocation of the servers to the queues with $\sum_{i = 1}^k \sum_{j = 1}^m L_i X_{ij} Y_{ij} \ge W. \hfill \diamond.$}

In the (PM) problem, we refer to the quantity $\sum_{i = 1}^k \sum_{j = 1}^m L_i X_{ij} Y_{ij}$ as the weight of the allocation.

{\thm\label{mw_hard} The MaxWeight problem (PM) is NP-complete.}

\begin{proof}
Please see Appendix~\ref{mw_hard_proof}.
\end{proof}

\section{The Batch-and-allocate Algorithm}
\label{sec:ba_algo}

The computational hardness results in Section~\ref{sec:hardness} imply that unless P=NP, there does not exist a computationally efficient scheduling algorithm that guarantees throughput optimality under general arrival and channel conditions. On the other hand, the user-experienced quality of service is crucially dependent upon a good delay performance. Hence we focus on designing a computationally tractable algorithm that gives a good delay performance under a restricted class of arrival and channel processes, namely, i.i.d. arrivals and channels with a bounded support, as specified in Section~\ref{sec:sysmodel}. We call this algorithm the Batch-and-allocate (BA) algorithm. We first define the Selective-allocate (SA) algorithm that is used as a ``black-box'' in the BA algorithm.

\smallskip
\hrule
\smallskip

\noindent \textbf{Selective-allocate (SA) algorithm:}\\
\textbf{Input:}
\begin{enumerate}
	\item An integer $k \ge 1.$
	\item A bipartite graph $G(\U \cup \V, \E)$ with $|\V| \ge k|\U|.$ Let $\U = \{u_1, \dots, u_x\}$ and $\V = \{v_1, \dots, v_y\}.$
\end{enumerate}

\noindent \textbf{Steps:}
\begin{enumerate}
	\item\label{step:partition} Partition the nodes in the set $\{v_1, \dots, v_{kx}\}$ into disjoint subsets $\V_1, \dots, \V_x$ such that $\V_i = \{v_{(i - 1)k + 1}, \dots, v_{ik}\}.$ Let $\V' := \{\V_1, \dots, \V_x\}.$
	\item Construct a new graph $H(\U \cup \V', \E')$ where an edge $(u_i, \V_j)$ is present in $\E'$ if the node $u_i$ is connected to every node in the set $\V_j$ in the original graph $G.$
	\item Find a largest cardinality matching $\M$ in the graph $H,$ breaking ties arbitrarily.
\end{enumerate}

\noindent \textbf{Output:} The matching $\M.$ \hfill $\diamond$

\smallskip
\hrule
\smallskip

The SA algorithm groups the nodes in the set $\V$ into sets of size $k$ each, and matches each such group $\V_i$ to that node $u_j \in \U$ that is connected to \emph{each} node in the group $\V_i.$ One can think of each node in the set $\U$ as a queue, each node in the set $\V$ as a server, and the presence of an edge signifies that the server can serve the given queue.
\begin{figure}[hbtp]
	\resizebox{3in}{!}{\input{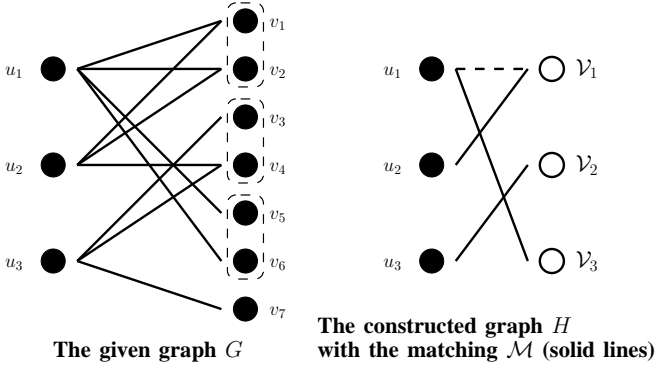}}
	\caption{SA algorithm - example}
	\label{fig:sa_example}
\end{figure}
An example of the SA algorithm for the case $k = 2$ is shown in Figure~\ref{fig:sa_example}. Here the solid edges in the graph $H$ represent the matching $\M.$ We write $\M = SA(k, G)$ for the output of the SA algorithm.
\smallskip
\hrule
\smallskip

\noindent \textbf{Batch-and-allocate (BA) algorithm:}\\
\textbf{Input:}
\begin{enumerate}
	\item The vector of queue-lengths, $Q_1(t - 1), \dots, Q_n(t - 1).$
	\item The vector of arrivals, $A_1(t), \dots, A_n(t).$
	\item The channel realizations, $X_{ij}(t)$ for $i, j \in [n].$
\end{enumerate}

\noindent \textbf{Steps:}
	
	\begin{enumerate}
	
	\item\label{step:housekeeping} Calculate $\hat{Q}(t - 1) := \max\limits_{1 \le i \le n} Q_i(t - 1).$ If $X_{ij}(t) < K$ for some pair $(i, j),$ then set $X_{ij}(t) = 0$ for that pair and use this value of $X_{ij}(t)$ throughout the rest of the algorithm.
	
	\item\label{step:partition_queues} For $1 \le r \le m_0,$ define
	\begin{eqnarray*}
	\D_r &:=& \{i \in [n] : \hat{Q}(t - 1) + (r - 1)K + 1 \le\\
	&& \; Q_i(t - 1) + A_i(t) \le \hat{Q}(t - 1) + rK\}
	\end{eqnarray*}
	to be the set of queue-indices $i$ such that the queue $Q_i$ needs to be allocated exactly $r$ servers to ensure $Q_i(t) \le \hat{Q}(t - 1).$	
	Let
	\[\D_0 = \{i \in [n]: Q_i(t - 1) + A_i(t) = \hat{Q}(t - 1)\}\]
	be the set of queue-indices $i$ such that after arrivals, the queue-length of $Q_i$ is the maximum queue-length at the end of the previous timeslot. We allocate servers to only some of the queues in the sets $\D_i, 0 \le i \le m_0.$ Let $d_i = |\D_i|.$
	
	\item\label{step:define_a} Let $0 \le a \le m_0 + 1$ be the smallest integer such that $\sum_{i = a}^{m_0} i d_i \le n.$ Here $a = m_0 + 1$ implies that the previous summation is vacuous (equal to~$0$), i.e., $m_0 d_{m_0} > n.$ Let $n' := n - \sum_{i = a}^{m_0} i d_i.$
	
	\item\label{step:partition_servers} \textbf{Case~$a \le m_0:$} Let $c \in \{a, a + 1, \dots, m_0\}$ be the largest integer such that $d_c + d_{c + 1} + \dots + d_{m_0} \ge n' / 2.$ For each $i \in \{c + 1, c + 2, \dots, m_0\}$ define the set of servers $\T_i$ satisfying
	\[|\T_i| = (i + 1)d_i + \frac{n'}{2(m_0 - a + 1)}.\]
	For each $i \in \{a, a + 1, \dots, c - 1\},$ define the set of servers $\T_i$ satisfying
	\[|\T_i| = id_i + \frac{n'}{2(m_0 - a + 1)}.\]
	Define $n'' = n' - (d_{c + 1} + d_{c + 2} + \dots + d_{m_0}).$	Define the set of servers $\T_c$ satisfying
	\[|\T_c| = (c + 1)n'' + c(d_c - n'') + \frac{n'}{2(m_0 - a + 1)}.\]
	Ensure that the servers in $\T_i$ are consecutively numbered for all $i \in \{a, a + 1, \dots, m_0\}.$

	\textbf{Case~$a = m_0 + 1:$} Define the set of servers $\T_{m_0} = \ess,$ the set of all the servers.
	
	\item\label{step:allocate} \textbf{Allocating servers to queues:}
	
	\textbf{Case~$a \le m_0:$}
	
		\begin{enumerate}
			\item[a)] For every $i \in \{c + 1, c + 2, \dots, m_0\},$ let $G_i$ be the restriction of the graph $G(\Q \cup \ess, \E)$ where the set of queues is restricted to indices in $\D_i,$ and the set of servers to $\T_i.$ Compute $\M_i = SA(i + 1, G_i).$ For every $i \in \{a, a + 1, \dots, c - 1\},$ let $G_i$ be the restriction of the graph $G(\Q \cup \ess, \E)$ where the set of queues is restricted to indices in $\D_i,$ and the set of servers to $\T_i.$ Compute $\M_i = SA(i, G_i).$ If $i = 0,$ compute $\M_0 = SA(1, G_0).$
			
			\item[b)] Let $\D_c' \subseteq \D_c$ be any subset satisfying $|\D_c'| = n'',$ and $\D_c'' = \D_c \setminus \D_c'.$ Let $\T_c' \subseteq \T_c$ be a subset satisfying $|\T_c'| = (c + 1) n'' + n'/(4(m_0 - a + 1))$ and $\T_c'' = \T_c \setminus \T_c'.$ Ensure that the servers in $\T_c', T_c''$ are consecutively numbered. Let $G_c'$ (resp. $G_c''$) be the restriction of $G$ where the set of queues is restricted to indices in $\D_c'$ (resp. $G_c''$), and the set of servers to $\T_c'$ (resp. $T_c''$).
				Compute $\M_c' = SA(c + 1, G_c')$ and $\M_c'' = SA(c, G_c'').$ Let $\M_c = \M_c' \cup \M_c''.$		
		\end{enumerate}
		
		For $a \le i \le m_0,$ allocate the servers to the queues as dictated by $\M_i:$ if $(Q_x, \W_y) \in \M_i$ for some queue $Q_x$ with $x \in \D_i$ and a set of servers $\W_y,$ then allocate the servers in $\W_y$ to $Q_x,$ etc., and accordingly define the allocation random variables $Y_{ij}(t).$
		
	\textbf{Case~$a = m_0 + 1:$} Let $G_{m_0}$ be the restriction of the graph $G(\Q \cup \ess, \E)$ where the set of queues is restricted to  indices in $\D_{m_0},$ and the set of servers to $\T_{m_0} = \ess.$ Compute $\M_{m_0} = SA(n / m_0, G_{m_0}).$ Allocate the servers to the queues as dictated by $\M_{m_0}.$

	\item\label{step:update} Update the queue-lengths to account for service as per Equation~\eqref{eq:queue_evol}.
	
\end{enumerate}

\noindent \textbf{Output:}
\begin{enumerate}
	\item The allocations, $Y_{ij}(t)$ for $i, j \in [n].$
	\item The final queue-lengths, $Q_i(t).$
\end{enumerate}

\smallskip
\hrule
\smallskip

Informally, the algorithm tries to reduce the queue-length of each of the queues after arrivals, to the maximum queue-length before arrivals. In order to limit the number of search possibilities, the algorithm only considers channels that have the maximum rate = $K.$ The algorithm groups the queues into disjoint sets such that the queues in each group require the same number of servers to attain a queue-length less that or equal to the maximum queue-length at the end of the previous timeslot. It then determines the number of servers to allocate to the queues in each group, which is somewhat more than the bare-minimum required number of servers to reduce each queue-length to the desired value. It assigns subsets of consecutively-numbered servers to each group of queues. The SA algorithm is used to make assignment decisions within each set of queues and the respective group of servers.

Some features of the algorithm are: (i)~This is a real-time algorithm; it does not need to know the statistical system parameters (e.g., the probabilities) in order to be implemented. (ii)~This algorithm results in a strictly positive value of the rate function (Theorem~\ref{ba_rate_func}). (iii)~This algorithm can be implemented in polynomial time (Theorem~\ref{ba_complexity}).

In order to limit complexity, the algorithm treats the smaller channel-rates as $0.$ In spite of this ``wastage,'' the algorithm gives a good small-queue performance (Theorem~\ref{ba_rate_func}). So the message is: for good delay performance, even under the single-carrier constraint, \emph{it is enough to focus on the highest-rate channels alone.} We first establish an important property of the SA algorithm.

{\lem\label{sa_match} Consider a graph $G(\U \cup \V, \E)$ with $|\V| = r \ge k|\U|.$ Suppose that for any pair of nodes $u \in \U, v \in V,$ the edge $(u, v)$ is present in $\E$ with probability $q,$ independently of all other random variables. Let $\M = SA(k, G).$ Then for $r$ large enough, $\probb(|\M| < |\U|) \le 3{\lfloor r/k \rfloor}(1 - q^k)^{\lfloor r/k \rfloor}.$}

\begin{proof}
Please see Appendix~\ref{sa_match_proof}.
\end{proof}

Note that the RHS of the above expression tends to $0$ as $r\to \infty$ for a fixed $k.$ Now our objective is to show that under the BA algorithm, in every timeslot, the probability that the maximum queue-length in the system increases is ``small'' for $n$ large. Define $m_0 := \lceil M/K \rceil.$

{\lem\label{ba_upcross} Fix any $\epsilon \in (0, \alpha / (2Mm_0)).$ Define the set $\B_\epsilon$ of probability measures ``near'' the distribution of the arrival process, as
\[
\B_\epsilon := \{[x_0, \dots, x_M] \in \Delta_{M + 1} : \ |x_i - p_i| < \epsilon \; \forall \ 0 \le i \le M\}.
\]
\[\textnormal{For $\epsilon \in \real_+,$ define} \quad \tau(\epsilon) := \inf_{\textbf{y} \in \Delta_{M + 1} \setminus \B_\epsilon} \sum_{i = 0}^M y_i \log \frac{y_i}{p_i}. \quad\]
Here $\tau: \real_+ \to \real_+ \cup \{\infty\}.$ Fix any $\rho \in (0, 1).$ Then under the BA algorithm, for $n$ large enough, for any timeslot $t,$
{\small
\begin{eqnarray*}
\lefteqn{\probb\left(\hat{Q}(t + 1) > \hat{Q}(t) \right)}\\
&\le& e^{-n\rho\tau(\epsilon)} + 3m_0\left\lfloor\frac{n\alpha}{4m_0(m_0 + 1)}\right\rfloor (1 - q_K^{m_0})^{\left\lfloor\frac{n\alpha}{4m_0(m_0 + 1)}\right\rfloor}.
\end{eqnarray*}
}
}

\begin{proof}
Please see Appendix~\ref{ba_upcross_proof}.
\end{proof}

We now show that for $n$ large, the probability that in a constant number of timeslots, the maximum queue-length in the system \emph{decreases} is at least $1/2.$

{\lem\label{ba_downcross} Under the BA algorithm, for $n$ large, there exists a constant integer $k_0$ such that
\[\probb\left(\hat{Q}(t + k_0) < \hat{Q}(t) - 1 \left| \hat{Q}(t) > 0 \right. \right) \ge \frac{1}{2}.\]
Further, $k_0 = \left\lceil \frac{4}{\alpha} \right\rceil$ is a valid choice.
}

\begin{proof}
Please see Appendix~\ref{ba_downcross_proof}.
\end{proof}

As a result of Lemmas~\ref{ba_upcross} and~\ref{ba_downcross}, the maximum queue-length in the system has the following behavior:
\begin{enumerate}
	\item In a given timeslot, it increases with probability that is exponentially small in $n,$ and if it increases, the amount of increase is no more than $M,$ which is a constant independent of $n.$
	\item Over a constant number of timeslots, it decreases with at least a constant ($= 1/2$) probability.
\end{enumerate}
Thus, it is reasonable to expect that the stationary distribution of the maximum queue-length is strongly concentrated near $0,$ which is formally established next.

{\thm \label{ba_rate_func} Under the BA algorithm, the stationary distribution of the maximum queue-length in the system obeys
\begin{eqnarray*}
\lefteqn{\liminf_{n \to \infty} \frac{-1}{n} \log\probb\left(\max_{1 \le i \le n} Q_i(t) > b\right)}\\
&& \ge \frac{b + 1}{M} \min\left(\tau(\epsilon), \frac{\alpha}{4m_0(m_0 + 1)} \log \frac{1}{1 - q_K^{m_0}}\right) > 0.
\end{eqnarray*}
}

\begin{proof}
Please see Appendix~\ref{ba_rate_func_proof}.
\end{proof}

Thus the proposed BA algorithm results in a strictly positive value of the rate function. Next we analyze its complexity.

{\thm\label{ba_complexity} The BA algorithm can be implemented in $\order(n^{2.5})$ computations per timeslot.}

\begin{proof}
Please see Appendix~\ref{ba_complexity_proof}.
\end{proof}

We conclude this section by showing that there is a finite upper bound on the rate-function \emph{under any algorithm.} The purpose is to establish that in the multi-queue multi-server setup considered in this paper, the probability of the overflow event decays like $e^{-n}$ at best; not like $e^{-n^2}$ or $e^{-n \log n},$ etc.

{\thm\label{rf_upper_bound} Fix $\theta \in (0, M/K - 1).$ Define $\C_\theta = \{\textbf{x} \in \Delta_{M + 1} : \sum_{i = 0}^M ix_i \ge K(1 + \theta)\}$ and
\[\xi(\theta) = \inf_{\textbf{y} \in \Delta_{M + 1} \setminus \C_\theta} \sum_{i = 0}^M y_i \log \frac{y_i}{p_i}.\]
Then under any algorithm for allocating servers to the queues,
\[\liminf_{n \to \infty} \frac{-1}{n} \log\probb\left(\max_{1 \le i \le n} Q_i(t) > b\right) \le \left\lceil \frac{b + 1}{\theta}\right\rceil \xi(\theta).\]
}

\begin{proof}
Please see Appendix~\ref{rf_upper_bound_proof}.
\end{proof}

Thus there is at most a constant-factor gap from optimality for the rate function under the BA algorithm.

\section{Extensions}
\label{sec:ext}

The BA algorithm presented in Section~\ref{sec:ba_algo} can be easily extended to a variety of cases of interest.

(i)~\textbf{Unequal number of queues and servers:} This case is of practical importance, because in typical uplink wireless systems, the number of active users is smaller than the number of orthogonal frequency sub-bands. The BA algorithm can be easily modified to utilize this ``extra'' service capacity, as follows. Suppose we have a system with $n$ users and $rn$ frequency sub-bands (servers) for some $r \ge 1.$ We refer to $r$ as the over-provision factor. In the step~\ref{step:partition_servers} of the BA algorithm, we give $r$ times as many servers to each group of queues $\D_i$ compared to the case of $n$ queues and $n$ servers. As a result, the rate-function lower bound of Theorem~\ref{ba_rate_func} scales up by a factor of $r.$ Formally, under the BA algorithm, the stationary distribution of the maximum queue-length in the system obeys
{\small
	\begin{eqnarray*}
	\lefteqn{\liminf_{n \to \infty} \frac{-1}{n} \log\probb\left(\max_{1 \le i \le n} Q_i(t) > b\right)}\\
	&& \ge \frac{r(b + 1)}{M} \min\left(\tau(\epsilon), \frac{\alpha}{4m_0(m_0 + 1)} \log \frac{1}{1 - q_K^{m_0}}\right) > 0.
	\end{eqnarray*}
	}
\noindent We omit the proof details.
	
(ii)~\textbf{Different priorities to queues:} The BA algorithm can be used in the case where the queues have different priorities. In this set up, we are interested in minimizing the probability of the event $\{\max_{i \in [n]} a_i Q_i(t) > b\}$ where $0 < a_{\min{}} \le a_i \le 1$ are given numbers. The BA algorithm instead operates on the ``effective'' queue-lengths, namely, $a_i Q_i(t),$ to yield rate-function results similar to Theorem~\ref{ba_rate_func}.

\section{Simulation Results}
\label{sec:sims}

We now analyze the performance of the proposed Batch-and-allocate (BA) algorithm through simulations. The goals are threefold: (i)~The rate-function results for the BA algorithm are asymptotic, i.e., as the number of users ($n$) and the number of sub-bands tend to infinity. We want to understand how large $n$ needs to be, to get a good small-buffer performance. (ii)~We want to understand the (good) impact of having more frequency sub-bands than the number of users, which is typically the case in today's wireless uplink systems. (iii)~We want to compare the BA algorithm's performance to an OFDMA-based greedy algorithm in~\cite{BodShaYinSri_infocom_10} that operates in the absence of the single-carrier constraint, in order to quantify the performance loss due to the single-carrier constraint. In the simulations, we run the OFDMA-based algorithm with as many servers as the users (i.e., over-provision factor, $r = 1$).

For simulation purpose, we arbitrarily assume an arrival process distribution of the form $(x + 1) e^{-x}$ on a bounded support $\{0, 1, \dots, 5\},$ normalized. We assume that the channel-rates are either $0$ or $2$ packets per timeslot. Thus $M = 5$ and $K = 2$ in the paper's notation. We refer to the quantity $\sum_{i = 1}^M p_i \left\lceil \frac{i}{K} \right\rceil$ as the effective load. In our case, the effective load is about $62\%.$ We vary the channel ON probability, $q,$ from $0.7$ to $0.9,$ and plot the empirical probability of buffer overflow v/s buffer-size, averaged over $10^6$ timeslots.

\begin{figure}[hbtp]
	\centering
	\includegraphics[width=4in]{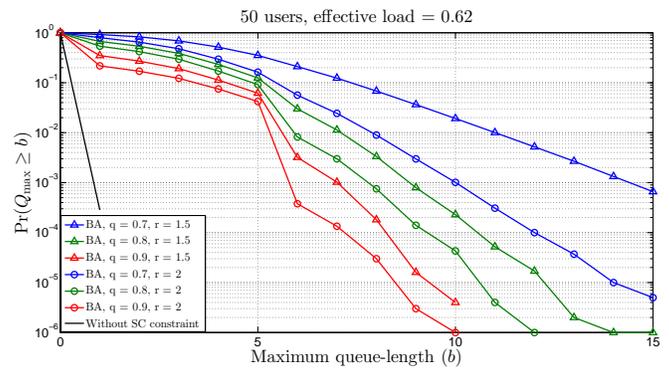}
	\caption{Performance of the BA algorithm}
	\label{fig:50users_effload_62pct}
\end{figure}

The results are presented in Figure~\ref{fig:50users_effload_62pct}. As we can see, the presence of the single-carrier constraint significantly degrades the small-buffer performance: the buffer overflow probabilities in the absence of the single carrier constraint are substantially lower than otherwise. We see that the buffer overflow probability \emph{decreases} with increasing system-size, as expected: the overflow probability is exponentially small in the system-size. We also see that changing the over-provisioning factor from $1.5$ to $2$ provides some performance boost. This confirms that the BA algorithm can seamlessly utilize more frequency sub-bands. Most interestingly, the asymptotic rate-function results for the BA algorithm already manifest themselves to give a good small-buffer performance at $n = 50.$ We have seen a comparable performance for the case $n = 40.$ Thus, the proposed BA algorithm yields a good small-queue performance at realistic system-sizes.

\section{Conclusions}
\label{sec:concl}

We considered the problem of user-scheduling in the wireless uplink networks. The distinguishing feature that makes this problem harder than the OFDM downlink scheduling problem is the presence of the single-carrier constraint. We showed that under the single-carrier constraint, the MaxWeight problem and the packet-draining problem are NP-complete. We presented the Batch-and-allocate algorithm that has polynomial complexity per timeslot, and a good small-queue performance for a class of bounded arrival and channel processes. The algorithm is robust to changes in the system-model. The results were validated through analysis and simulations.

\section*{Acknowledgments}

The authors would like to thank Nilesh Khude and Saurabh Tavildar for helpful discussions.

\appendices

\section{Proof of Theorem~\ref{pd_hard}}
\label{pd_hard_proof}

The problem (PD) clearly belongs to the class NP: a certificate is an allocation of the servers to the queues that serves a total of at least $W$ packets from the queues. In order to show that it is NP-complete, we use a reduction to the Hamiltonian path problem, which is NP-complete (\cite{kleinberg_tardos}, Ch.~8). The Hamiltonian path problem asks: given a directed graph $G,$ does it contain a (directed) path, starting and ending at any node, that visits every node exactly once?

\noindent \textbf{Reduction:}\\
\noindent Given a directed graph $G(\V, \E)$ with $|\V| = n,$ we construct a directed bipartite graph $G'(\V_\ell \cup \V_r, \E')$ as follows: for every node $v_i \in \V,$ define two nodes $v_{\ell, i} \in \V_\ell$ and $v_{r, i} \in \V_r.$ Connect $v_{\ell, i}$ to $v_{r, i}$ via a directed edge. If a directed edge $(v_i, v_j)$ exists in $\E,$ then introduce a directed edge from $v_{r, i}$ to $v_{\ell, j}.$ That is, all the incoming edges to $v_i$ are connected to $v_{\ell, i}$ and all the outgoing edges are connected to $v_{r, i}.$ One can easily show that the graph $G$ has a Hamiltonian cycle iff $G'$ has; we omit the proof. We call the graph $G'$ the bipartite version of $G.$

Define $T = 2n(n + 1)(n + 2).$ Consider an instance of the problem (PD) with $2n$ queues, each with $2nT + (2n - 1)(n + 2)$ packets, and $2nT + (2n - 1)(n + 2)$ servers.
The servers are grouped in $4n - 1$ sets: $2n$ sets of $T$ servers each, and $2n - 1$ sets of $n + 2$ servers each. Let the sets of $T$ servers be called $A_{\ell, 1}, \dots, A_{\ell, n}, A_{r, 1}, \dots, A_{r, n},$ and the sets of $n + 2$ servers be called $B_1, \dots, B_{2n - 1}.$ The servers within a set are consecutively indexed. We use the symbol $C < D$ to denote that the servers in the set $C$ have lower indices than those in the set $D.$ We order the servers such that
\begin{eqnarray*}
A_{\ell, 1} < B_1 < A_{r, 1} < B_2 < A_{\ell, 2} < B_3 < A_{r_2} \\
< B_4 < A_{\ell, 3} < \dots < B_{2n - 1} < A_{r, n}.
\end{eqnarray*}

Let the set of $2n$ queues be $Q_{\ell, 1}, \dots, Q_{\ell, n}, Q_{r, 1}, \dots, Q_{r, n}.$ Let $X(Q_{\ell, i}, S_j)$ denote the number of packets that the server $S_j$ can serve from the queue $Q_{\ell, i},$ and similarly for $Q_{r, i}.$

Fix a queue $Q_{\ell, i}.$ Note that a server in the set $B_x$ has the index $xT + (x - 1)(n + 2) + j$ for $j \in [n + 2].$

\begin{enumerate}
	\item\label{step:self} For every $x \in [n]$ and for each server $S_j \in A_{\ell, x},$ define $X(Q_{\ell, i}, S_j) = 1.$ For every $x \in [n]$ and for each server $S_j \in A_{r, x},$ define $X(Q_{\ell, i}, S_j) = 0.$
	\item\label{step:odd} Fix $x \in [2n - 1],$ $x$ odd. For a server in $B_x$ with index $xT + (x - 1)(n + 2) + j,$ define $X(Q_{\ell, i}, S_{xT + (x - 1)(n + 2) + j}) = i + 1$ if $i = j,$ and $0$ otherwise.
	\item\label{step:even} Let $v_{r, g_1}, v_{r, g_2}, \dots$ be the nodes that have an outgoing edge to the node $v_{\ell, i},$ with $g_1 > g_2 > \dots.$ Fix $x \in [2n - 1],$ $x$ even, and a server in $B_x$ with index $xT + (x - 1)(n + 2) + g_j.$ Define $X(Q_{\ell, i}, S_{xT + (x - 1)(n + 2) + g_1}) = n + 1 - g_1$ and $X(Q_{\ell, i}, S_{xT + (x - 1)(n + 2) + g_j}) = n + 1 - (g_{j - 1} - g_j)$ for $j > 1.$ Define $X(Q_{\ell, i}, S_{xT + (x - 1)(n + 2) + j}) = 0$ for all other values of $j \in [n + 2].$	
\end{enumerate}

Perform the same construction for a queue $Q_{r, i}$ with $A_{r, x}$ replacing $A_{\ell, x}$ and vice-versa in step~\ref{step:self}, and the words ``even'' and ``odd'' replacing each other in steps~\ref{step:odd} and~\ref{step:even}. Define the number $W = 2nT + (2n - 1)(n + 2).$

Here the steps~\ref{step:self} and~\ref{step:odd} are generic and apply to any graph $G,$ while step~\ref{step:even} is dependent upon the graph structure. We now establish some basic properties of the possible server allocations for the above construction under the single carrier constraint.

\medskip

\noindent \textbf{Property~1:} If a queue, say $Q_{\ell, i}$ is not allocated any server from the set $\cup_{x \in [n]} A_{\ell, x},$ then the maximum total number of packets that can be served from the queues is less than $2nT.$

\noindent \textbf{Proof:} A server in a set $A_{\ell, x}$ can serve at most $1$ packet. Further, that packet must be from a queue labeled $Q_{\ell, x'}$. Thus the maximum number of packets served by all the servers in the set $A_{\ell, x}$ is $T.$ Suppose that for all $x \in [n],$ the servers in the sets $A_{\ell, x}$ serve $1$ packet each. Since by hypothesis at most $n - 1$ queues in $\{Q_{\ell, 1}, \dots, Q_{\ell, n}\}$ can be allocated a server in $A_{\ell, x},$ by the pigeonhole principle, at least one queue $Q_{\ell, j}$ must be served by servers in $A_{\ell, x}$ and $A_{\ell, x + r}$ for some $x,$ some $r \ge 1.$ Consequently, as a result of the single carrier constraint, all the servers in $A_{r, x}$ must be allocated to $Q_{\ell, j},$ each serving $0$ packets from $Q_{\ell, j}.$ Thus the maximum number of packets that can be served by the servers in $A_{r, \cdot}$ is $(n - 1)T.$ The total number of packets that can be served by the servers in $A_{\ell, \cdot}$ is at most $nT.$

For $x \in [2n - 1],$ the maximum number of packets that can be served by a server in $B_x$ is $n + 1.$ The total number of servers in any set $B_x$ is $n + 2.$ Thus the total number of packets that can be served by \emph{all} the servers in $\cup_x B_x$ is $(2n - 1)(n + 1)(n + 2).$ Since $T = 2n(n + 1)(n + 2),$ we have $(n - 1)T + nT + (2n - 1)(n + 1)(n + 2) < 2nT.$ \hfill $\spadesuit$

By symmetry, the above property also holds for a queue $Q_{r, i}.$ Thus, any allocation that serves $W = 2nT + (2n - 1)(n + 2)$ packets must allocate server(s) from $\cup_{x \in [n]} A_{\ell, x}$ (resp. $\cup_{x \in [n]} A_{r, x})$ to each queue $Q_{\ell, i}$ (resp. $Q_{r, i}$).

\medskip

\noindent \textbf{Property~2:} Exactly one of the following statements is true:
\begin{enumerate}
	\item\label{stmt:perm} There exists a permutation $\sigma$ (resp. $\pi$) of $[n]$ such that all the servers in $A_{\ell, i}$ (resp. $A_{r, i}$) are allocated to $Q_{\ell, \sigma_i}$ (resp. $Q_{r, \pi_i}$).
	\item\label{stmt:unsat} The allocation serves a total of $W' < 2nT$ packets from all the queues.
\end{enumerate}

\noindent \textbf{Proof:} If statement~\ref{stmt:perm} holds, then evidently a total of $2nT$ or more packets are served, so statement~\ref{stmt:unsat} cannot hold. If statement~\ref{stmt:perm} does not hold, then WLG suppose a queue $Q_{\ell, i}$ is allocated servers from $A_{\ell, x}$ and $A_{\ell, x + r}$ for some $x,$ some $r \ge 1.$ As before, the servers in $A_{r, x}$ are allocated to $Q_{\ell, i},$ serving $0$ packets. Hence, as established in the proof of Property~1, the allocation serves a total of $W' < 2nT$ packets, thus statement~\ref{stmt:unsat} holds. \hfill $\spadesuit$

Thus, if an allocation serves $2nT$ or more packets, then for every $x \in [2n - 1],$ the set of servers in $B_x$ serve at most $2$ queues, and the queues (if two) are of the form $(Q_{\ell, i}, Q_{r, j}).$

\medskip

\noindent \textbf{Property~3:} Let an allocation serve a total of at least $2nT$ packets from all the queues. If all the servers in a set $B_x$ are allocated to the same queue, say $Q_{\ell, i},$ then the total number of packets served by the servers in $B_x$ is at most $n + 1.$

\noindent \textbf{Proof:} If $x$ is odd, then exactly $1$ server from $B_x$ can serve a nonzero number of packets from $Q_{\ell, i},$ and that number equals $i + 1.$ If $x$ is even, then the server $B_x$ can serve at most $n + 1 - 1 \le n + 1$ packets from $Q_{\ell, i}.$ \hfill $\spadesuit$

An allocation of servers to the queues is said to be \emph{normal} if there exists a permutation $\sigma$ (resp. $\pi$) of $[n]$ such that all the servers in $A_{\ell, i}$ (resp. $A_{r, i}$) are allocated to $Q_{\ell, \sigma_i}$ (resp. $Q_{r, \pi_i}$).

\medskip

\noindent \textbf{Property~4:} Fix $x \in [n],$ $x$ odd. Under a normal allocation, let the servers in $B_x$ serve two queues $(Q_{\ell, i}, Q_{r, j}).$ If there exists a directed edge $(v_{\ell, i}, v_{r, j}) \in \E',$ then the servers in $B_x$ serve a total of at most $n + 2$ packets, else, serve at most $n + 1$ packets.

\noindent \textbf{Proof:} Suppose the servers in $B_x$ serve two queues $(Q_{\ell, i}, Q_{r, j})$ with $(v_{\ell, i}, v_{r, j}) \in \E'.$ There is exactly one server $S_t$ in $B_x$ that serves $Q_{\ell, i}$ at a nonzero rate of $i + 1$ packets. If this server $S_t$ is not allocated to $Q_{\ell, i},$ then the number of packets served from $Q_{r_j}$ is at most $n + 1$ by Property~3: the number of packets served from $Q_{r, j}$ cannot be more than if all the servers in $B_x$ are allocated to $Q_{r, j}.$

If $S_t$ is allocated to $Q_{\ell, i},$ then because $x$ is odd and the allocation is normal, the servers in $B_x$ with indices less than $t$ are allocated to $S_t.$ The maximum number of packets that can be served from $Q_{r, j}$ by allocating to it all the servers in $B_x$ with indices higher than $t$ is $n - i + 1,$ implying a total of $n + 2$ packets at most.

If there does not exist a directed edge $(v_{\ell, i}, v_{r, j})$ in $\E',$ then, even after allocating $S_t$ to $Q_{\ell, i}$ and all the servers in $B_x$ with indices higher than $t$ to $Q_{r, j},$ the maximum number of packets served from $Q_{r, j}$ is at most $n - z + 1$ for $z > i,$ implying a total of at most $n + 1$ packets. \hfill $\spadesuit$

If the allocation of servers in $B_x$ to the queues $(Q_{\ell, i}, Q_{r, j})$ serves a total of $n + 2$ packets, we call it a drain-maximizing allocation for $B_x.$

A similar statement to Property~4 can be proved for $B_x$ for even $x,$ and an edge $(Q_{r, j}, Q_{\ell, i}) \in \E'.$ We are now in a position to prove that a Hamiltonian path exists in $G'$ if and only if there exists an allocation of servers to the queues that serves at least $W = 2nT + (n + 2)(2n - 1)$ packets. First suppose there exists an allocation that serves at least $W$ packets. Then it must be normal, and for every $B_x,$ it serves exactly $2$ queues, one from $\{Q_{\ell, 1}, \dots, Q_{\ell, n}\}$ and the other from $\{Q_{r, 1}, \dots, Q_{r, n}\},$ and the \emph{same} queues $Q_{\ell, i}$ and $Q_{r, j}$ that are served by the adjacent servers in sets $A_{\ell, \cdot}$ and $A_{r, \cdot}.$ Thus the queues $Q_{\ell, \sigma_1}, Q_{r, \pi_1}, Q_{\ell, \sigma_2}, Q_{r, \pi_2}, \dots, Q_{\ell, \sigma_n}, Q_{r, \pi_n}$ are served in order in consecutive server blocks. Consider the path $v_{\ell, \sigma_1} \to v_{r, \pi_1} \to v_{\ell, \sigma_2} \to v_{r, \pi_2} \to \dots \to v_{\ell, \sigma_n} \to v_{r, \pi_n}.$ This is a valid \emph{path} in the graph $G'$ (Property~4) and because $\sigma, \pi$ are permutations, it visits every node exactly once. Therefore it is a Hamiltonian path.

Next suppose that there is a Hamiltonian path in $G',$ WLG call it $v_{\ell, \sigma_1} \to v_{r, \pi_1} \to v_{\ell, \sigma_2} \to v_{r, \pi_2} \to \dots \to v_{\ell, \sigma_n} \to v_{r, \pi_n}.$ Then allocating to the queue $Q_{\ell, \sigma_i}$ the servers in $A_{\ell, i},$ to the queue $Q_{r, j}$ the servers in $\pi_j,$ and the drain-maximizing allocations for each $B_x$ (which is possible because of Property~4), we get an allocation that serves exactly $W = 2nT + (n + 2)(2n - 1)$ packets. This completes the reduction. Since $T = \order(n^3),$ this is a polynomial-time reduction. Therefore the problem (PD) is NP-complete.

\section{Proof of Theorem~\ref{mw_hard}}
\label{mw_hard_proof}

The problem (PM) clearly belongs to the class NP: a certificate is an allocation of the servers to the queues that has a weight of at least $W.$ To show that it is NP-complete, we use the same reduction to the Hamiltonian path problem as before, we consider each queue to be of length = $1$ packet, and ask the question whether $W = 2nT + (2n - 1)(n + 2)$ units of total service can be offered, which translates to a schedule-weight of $W.$ We omit the details.

\section{Proof of Lemma~\ref{sa_match}}
\label{sa_match_proof}

Let $z = {\lfloor r/k \rfloor}.$ Adding dummy nodes if necessary to the set $\U,$ and removing some nodes if necessary from the set $\V,$ we construct a graph $G'(\U' \cup V', \E')$ where $|\V'| = kz$ and $|\U'| = z.$ For a pair of nodes $u' \in \U'$ and $v' \in \V',$
\begin{enumerate}
	\item If $u' \in \U,$ then for any $v' \in \V' \subseteq \V,$ $(u', v') \in \E'$ if and only if $(u', v') \in \E.$
	\item If $u' \notin \U,$ then for any $v' \in \V',$ the edge $(u', v') \in \E'$ with probability $q,$ independently of all other random variables.
\end{enumerate}

Group the nodes in the set $\V'$ as described in the SA algorithm, to get a bipartite graph $G''(\U' \cup \V'', \E'')$ where $\V''$ is the set of groups of nodes in $\V',$ and nodes $u' \in \U', v'' \in \V''$ are connected by an edge in $\E''$ if the node $u'$ is connected to every node in the group $\V''.$ Thus between any pair of nodes in $\U' \times \V'',$ an edge exists with probability $q^k.$

For $z$ large enough, the graph $G''$ has a perfect matching $\M''$ with probability at least $1 - 3z(1 - q^k)^z$ (\cite{BodShaYinSri_sig_09}, Lemma~1). Removing the ``dummy'' nodes that were added to get the set $\U'$ from $\U,$ we get a matching $\M$ as the output of the SA algorithm with $|\U| = |\M|.$ That is, a perfect matching in the graph $G''$ (deterministically) yields a matching of cardinality $|\U|$ as the output of the SA algorithm. Therefore, for $r$ large enough, $\probb(|\M| < |\U|) \le 3{\lfloor r/k \rfloor}(1 - q^k)^{\lfloor r/k \rfloor}.$

\section{Proof of Lemma~\ref{ba_upcross}}
\label{ba_upcross_proof}

The proof proceeds in two steps: first we show that for large $n,$ with high probability, $a = 0$ holds in the step~\ref{step:define_a} of the BA algorithm. In the process, we show that the the number of ``excess servers'' $n'$ (step~\ref{step:define_a} of the BA algorithm) is at least $n\alpha/2$ with high probability. Next, under the condition $a = 0$ and $n' \ge n\alpha / 2,$ we show that the probability of $\{\hat{Q}(t + 1) > \hat{Q}(t)\}$ is small.

\noindent \textbf{Step~1:}\\
For $0 \le i \le M,$ let $p'_i := |\{k \in [n] : A_k(t + 1) = i\}| / n$ be the fraction of the $n$ queues that see exactly $i$ arrivals in the timeslot $t + 1.$ Let $\textbf{p'} = [p_0', p_1', \dots, p_M'].$ Choose any $\epsilon \in (0, \alpha / (2Mm_0)),$ say $\epsilon = \alpha / (4Mm_0).$ By Sanov's theorem (\cite{dembozeitouni}, Thm.~{2.1.10}), for any $\rho \in (0, 1),$ for $n$ large enough, $\probb(\textbf{p'} \notin \B_\epsilon) \le e^{-n\rho\tau(\epsilon)}.$ Since the set $\Delta_{M + 1} \setminus \B_\epsilon$ is compact and the function $g(\textbf{y}) = \sum_{i = 0}^M y_i \log (y_i / p_i)$ is lower semicontinuous (\cite{dembozeitouni}, Chapter~2, Exercise~{2.1.22}), the infimum in the definition of $\tau(\cdot)$ is achieved and is strictly positive ($\because g(\textbf{y}) = 0 \Leftrightarrow \textbf{y} = \textbf{p}, \textbf{p} \in \B_\epsilon$ and $g(\textbf{y}) \ge 0$ for all $\textbf{y}$). Thus $\tau(\epsilon) > 0,$ implying
\[\probb(|p_i - p'_i| < \epsilon, \forall \ i \in \{0, 1, \dots, M\}) \ge 1 - e^{-n\rho\tau(\epsilon)}.\] 

Let $\hat{Q}(t) = m.$ Define the set $\C_r := \{i \in [n] : (r - 1)K + 1 \le A_i(t + 1) \le rK\}.$ Since $Q_i(t) \le m$ for all $i,$ $\D_r \subseteq \bigcup_{i = r}^{m_0} \C_i.$ Hence,
\begin{eqnarray*}
|\D_r| &\le& |\C_r| + |C_{r + 1}| + \dots + |C_{m_0}|\\
&=& n(p'_{(r - 1)K + 1} + p'_{(r - 1)K + 2} + \dots + p'_M)
\end{eqnarray*}
implying
\begin{eqnarray*}
\sum_{r = 1}^{m_0} r |\D_r| = \sum_{r = 0}^{m_0} r |\D_r| &\le& n\sum_{i = 1}^{m_0} i \left(\sum_{j = (i - 1)K + 1}^M p'_j\right)\\
&=& n \sum_{i = 1}^M p'_i \left\lceil \frac{i}{K} \right\rceil\\
&\stackrel{(a)}{\le}& n \sum_{i = 1}^M (p_i + \epsilon) \left\lceil \frac{i}{K} \right\rceil\\ 
&\le& n(1 - \alpha) + n\epsilon M m_0,
\end{eqnarray*}
\noindent where the step~$(a)$ holds with probability at least $1 - e^{-n\rho\tau(\epsilon)}.$ Since $\epsilon < \alpha / (2Mm_0),$ we have $\sum_{r = 0}^{m_0} r |\D_r| \le n - n\alpha / 2,$ or $a = 0$ and $n' \ge n\alpha / 2$ in the step~\ref{step:define_a} of the BA algorithm, with probability at least $1 - e^{-n\rho\tau(\epsilon)}.$

\noindent \textbf{Step~2:}\\
We assume that $a = 0$ and $n' \ge n\alpha / 2$ in the step~\ref{step:define_a} of the BA algorithm. Consider the event $\E_i$ that each of the queues in the set $\D_i$ are allocated at least $i$ servers. If the event $\E_i$ occurs for every $i \in \{1, 2, \dots, m_0\},$ then the maximum queue-length at the end of timeslot $t + 1$ is at most $m.$ This event ($\E_i$) occurs if, in the server allocation step (step~\ref{step:allocate}) of the BA algorithm, the matching obeys $|\M_i| = |D_i|.$

Fix any $i \in \{1, 2, \dots, m_0\}.$ We have $|\T_i| \ge i |\D_i| + n'/(2(m_0 - a + 1)) \ge i|\D_i| + n'/(2(m_0 + 1)),$ and $|\T_i| / i \ge |\D_i| + n'/ (2m_0(m_0 + 1)) \ge |\D_i| + n\alpha / (4m_0(m_0 + 1)).$ Thus, from Lemma~\ref{sa_match},
{\small
\[\probb(|\M_i| = |\D_i|) \ge 1 - 3\left\lfloor\frac{n\alpha}{4m_0(m_0 + 1)}\right\rfloor (1 - q_K^{m_0})^{\left\lfloor\frac{n\alpha}{4m_0(m_0 + 1)}\right\rfloor}.\]
}

Hence, by the union bound,
\begin{eqnarray*}
\lefteqn{\probb(|\M_i| = |\D_i| \ \forall \ i \in [m_0])}\\
&\ge& 1 - 3m_0\left\lfloor\frac{n\alpha}{4m_0(m_0 + 1)}\right\rfloor (1 - q_K^{m_0})^{\left\lfloor\frac{n\alpha}{4m_0(m_0 + 1)}\right\rfloor}.
\end{eqnarray*}

Combining the results of steps~1 and~2 and once again using the union bound,
{\small
\begin{eqnarray*}
\lefteqn{\probb\left(\hat{Q}(t + 1) > \hat{Q}(t) \right)}\\
&\le& e^{-n\rho\tau(\epsilon)} + 3m_0\left\lfloor\frac{n\alpha}{4m_0(m_0 + 1)}\right\rfloor (1 - q_K^{m_0})^{\left\lfloor\frac{n\alpha}{4m_0(m_0 + 1)}\right\rfloor},
\end{eqnarray*}
}
\noindent completing the proof.

\section{Proof of Lemma~\ref{ba_downcross}}
\label{ba_downcross_proof}

Suppose at the end of timeslot $t,$ the maximum queue-length is $m$ and the number of queues at length $m$ is $x.$ Our objective is to show that at the end of timeslot $t + 1,$ with probability at least $1 - e^{-n\phi}$ for some $\phi > 0,$
\begin{enumerate}
	\item\label{prop:length} the maximum queue-length is at most $m,$ and
	\item\label{prop:size} the number of queues at the maximum is at most $(x - n\alpha/4)^+.$
\end{enumerate}
Since $x \le n,$ the properties~\ref{prop:length},~\ref{prop:size} and the union bound imply that with probability at least $1 - k_0 e^{-n\phi},$ at the end of $k_0 = \left\lceil\frac{4}{\alpha}\right\rceil$ timeslots, the maximum queue-length is at most $m - 1.$

First consider the case $x = n,$ i.e., all the queues in the system are equal in length. From Lemma~\ref{ba_upcross}, for $n$ large, the probability that $\hat{Q}(t + 1) \ge m$ is upper-bounded by $e^{-n\theta_1}$ for some $\theta_1 > 0,$ so the property~\ref{prop:length} is satisfied. Next, the BA algorithm allocates to the queues in the sets $\D_{c + 1}, \D_{c + 2}, \dots, \D_{m_0}$ \emph{one more server} than is necessary to bring their length to $m,$ and also for $n'' = n' - (d_{c + 1} + d_{c + 2} + \dots + d_{m_0})$ queues in $\D_c.$ Thus, at the end of timeslot $t + 1,$ the number of queues at length $m$ is at most $(n - n'/2)^+,$ and by the proof of Lemma~\ref{ba_upcross}, the probability of this event is at least $1 - e^{-n\theta_2}$ for some $\theta_2 > 0.$ Since $n' \ge n\alpha/2$ with probability at least $1 - e^{-n\theta_3}$ for some $\theta_3 > 0$ (from the proof of Lemma~\ref{ba_upcross}), if we choose $\phi = \min(\theta_1, \theta_2, \theta_3),$ then the property~\ref{prop:size} is satisfied for the case $x = n.$ The case $x < n$ is almost identical; we omit the details for the sake of brevity.

\section{Proof of Theorem~\ref{ba_rate_func}}
\label{ba_rate_func_proof}

The proof is almost identical to that of Theorem~5 in~\cite{BodShaYinSri_infocom_11}. In particular, Lemma~\ref{ba_downcross} shows that the maximum queue-length in the system decreases by at least $1$ (provided it is nonzero to begin with) over a constant number of timeslots, with probability at least $1/2.$ Lemma~\ref{ba_upcross} shows that in a given timeslot, it increases by at most $M,$ and the probability of this increase it at most $e^{-n\zeta}$ for some $\zeta = \zeta(\epsilon, \alpha, \rho) > 0,$ for $n$ large. Using the same stationary distribution bounding techniques as those in the proof of Theorem~5 in~\cite{BodShaYinSri_infocom_11}, we conclude that
\begin{eqnarray*}
\lefteqn{\liminf_{n \to \infty} \frac{-1}{n} \log\probb\left(\max_{1 \le i \le n} Q_i(t) > b\right)}\\
&\ge& \frac{b + 1}{M}\zeta(\epsilon, \alpha, \rho)\\
&=& \frac{b + 1}{M} \min\left(\rho\tau(\epsilon), \frac{\alpha}{4m_0(m_0 + 1)} \log \frac{1}{1 - q_K^{m_0}}\right) > 0,
\end{eqnarray*}
\noindent implying the desired result because $\rho < 1$ is arbitrary (formally, taking the limit of both sides as $\rho \to 1$).

\section{Proof of Theorem~\ref{ba_complexity}}
\label{ba_complexity_proof}

The steps~\ref{step:housekeeping} and~\ref{step:update} of the BA algorithm can be performed in $\order(n^2)$ computations each. The steps~\ref{step:partition_queues} and~\ref{step:partition_servers} can be performed in $\order(n)$ computations each. The step~\ref{step:define_a} can be performed in $\order(1)$ computations.

Step~\ref{step:allocate} requires finding largest cardinality matchings in bipartite graphs. Given a bipartite graph with $\order(n)$ nodes, the largest cardinality matching can be found in $\order(n^{2.5})$ computations~\cite{hk73}. In our case, we need to find largest cardinality matchings in bipartite graphs with $2n_1, 2n_2, \dots, 2n_w$ nodes respectively with $n_1 + n_2 + \dots + n_w = n.$ Hence the computational effort is $\order(n_1^{2.5} + n_2^{2.5} + \dots + n_w^{2.5}) = \order(n^{2.5}).$ Thus, the BA algorithm can be implemented in $\order(n^{2.5})$ computations per timeslot.

\section{Proof of Theorem~\ref{rf_upper_bound}}
\label{rf_upper_bound_proof}

Consider the following event that leads to overflow: fix $\theta \in (0, M/K - 1),$ and for $t_0 = \left\lceil \frac{b + 1}{\theta}\right \rceil$ timeslots up to and including the timeslot $0,$ the total number of arrivals to all the queues have an empirical mean $\ge nK(1 + \theta).$ That is, if $f_i(t) = \frac{1}{n} \sum_{j = 1}^n \ind\{A_j(t) = i\},$ then for $-t_0 < t \le 0,$ we have $\sum_{i = 0}^M i f_i(t) \ge K(1 + \theta).$ Since the system can serve at most $nK$ packets in a given timeslot, this event leads to an overflow at the end of timeslot $0$ \emph{under any algorithm.}

\noindent \textbf{Analyzing the probability of the event that leads to overflow:}
Fix any $\rho \in (0, 1).$ By Sanov's theorem (\cite{dembozeitouni}, Thm.~{2.1.10}), for any timeslot $t,$ the probability of the empirical mean of the arrivals exceeding $K(1 + \theta)$ is at least $e^{-n\rho\xi(\theta)}$ for $n$ large. Since $\theta < M/K - 1,$ the set $\Delta_{M + 1} \setminus \C_\theta$ is nonempty: $[0, 0, \dots, 0, 1] \in \Delta_{M + 1} \setminus \C_\theta.$ Hence, by the usual arguments of compactness and lower semicontinuity, the infimum in the definition of $\xi(\cdot)$ is achieved and is finite and strictly positive. By the independence of arrivals across timeslots, the probability of overflow event is thus at least $e^{-n\rho t_0\xi(\theta)},$ implying (because $\rho < 1$ is arbitrary)
\[\liminf_{n \to \infty} \frac{-1}{n} \log\probb\left(\max_{1 \le i \le n} Q_i(t) > b\right) \le \left\lceil \frac{b + 1}{\theta}\right\rceil \xi(\theta).\]

\end{document}